\documentstyle[aaspp4,epsf]{article}

\def\deg{$^{\circ}$}
\def\bs{$\Box ^{\prime\prime}$}
\def\as{$^{\prime\prime}$} 
\def\muas{$\mu$arcsec}

\def\bdm{\begin{displaymath}} 
\def\edm{\end{displaymath}} 
\def\beq{\begin{equation}} 
\def\eeq{\end{equation}} 
\def\bit{\begin{itemize}} 
\def\eit{\end{itemize}} 
\def\ben{\begin{enumerate}} 
\def\een{\end{enumerate}} 
\def\bfi{\begin{figure}[htb]} 
\def\efi{\end{figure}}

\def\msol{$\rm M_{\odot}$}
\def\mbh{$\rm M_{\bullet}$}
\def\ea{{\it et al.}}

\def\ha{$\rm H\alpha$}
\def\ss{\sc SIMsim\rm}
\def\bpic{$\beta$~Pic}
\def\uv{($u$,$v$)} 
\def\sinc{\rm sinc\it}

\def\plotone#1{\centering \leavevmode
\epsfxsize=\textwidth \epsfbox{#1}}

\begin{document}

\lefthead{B\"oker \& Allen}
\righthead{Imaging and Nulling with SIM}
\title{Imaging and Nulling with the Space Interferometer Mission}

\author{Torsten B\"oker\altaffilmark{1} \& Ronald J. Allen} 
\affil{Space Telescope Science Institute, 3700 San Martin Drive, 
Baltimore, MD 21218, U.S.A.}
\altaffiltext{1}{Affiliated with the Astrophysics Division, Space Science Department,
European Space Agency} 
\authoremail{boeker@stsci.edu, rjallen@stsci.edu}
\begin{abstract}
We present numerical simulations for a possible synthesis imaging mode of
the Space Interferometer Mission (SIM). We summarize the general techniques
that SIM offers to perform imaging of high surface brightness sources, 
and discuss their
strengths and weaknesses. We describe an interactive software
package that is used to provide realistic, photometrically correct 
estimates of SIM performance
for various classes of astronomical objects. In particular, we simulate the
cases of gaseous disks around black holes in the nuclei of galaxies, and
zodiacal dust disks around young stellar objects.
Regarding the first, we show that a Keplerian velocity gradient of the 
line-emitting gaseous disk --- and 
thus the mass of the putative black hole --- can be determined with SIM to 
unprecedented accuracy in about 5~hours of integration time for objects
with \ha\ surface brigthness comparable to the prototype M~87. 
Detections and observations of exo-zodiacal dust disks depend critically
on the disk properties and the nulling capabilities of SIM. Systems with
similar disk size and at least one tenth of the dust content of \bpic\
can be detected by SIM at distances
between 100~pc and a few kpc, if a nulling efficiency of $10^{-4}$ is
achieved. Possible inner clear regions indicative of the presence of
massive planets can also be detected and imaged. 
On the other hand, exo-zodiacal disks with properties more similar to
the solar system will not be found in reasonable integration times 
with SIM.
\end{abstract}
\section{Introduction}\label{intro}
For certain areas of current astronomical research, such as the detection and
mass determination of black holes in galactic nuclei, spatial resolution is
the main limitation. For others, e.g. the case of extrasolar planets,
the challenge is to detect very faint signals in the presence of an
overwhelmingly bright star. Interferometry from space has the
potential to overcome both of these obstacles: it has no fundamental limit to
spatial resolution, and interferometric nulling is by far the most powerful 
method to suppress unwanted light from a bright point source.

The Space Interferometer Mission (SIM) will be the first mission to demonstrate
the feasibility of this new and exciting technology. Scheduled for launch in 2005,
SIM is the first new mission of NASA's ORIGINS program after SIRTF. It is designed
primarily for astrometry at the \muas -level. However, SIM may also have
significant capabilities for synthesis imaging. In
terms of spatial resolution, SIM with a maximum baseline of 12~m promises
at least a factor of six improvement over the Hubble Space Telescope (HST).

In this paper, we discuss the potential of SIM to significantly improve
upon current observational limits in two specific astronomical cases, namely
line-emitting gas disks rotating around black holes in galactic nuclei, and 
continuum zodiacal
dust disks around nearby stars. In the next section, we will describe
how SIM performs synthesis imaging, and discuss some aspects
and limitations that will affect the observations. 
Section~\ref{simsim} gives an overview of the capabilities of the code 
which we use to simulate 
various classes of model sources. Input models and first results of 
simulations for the two examples mentioned above are described in 
Sections~\ref{bh} and \ref{exo}. We discuss and summarize our findings 
in Section~\ref{sum}.
Appendices describe the basics of interferometric imaging, and the
simulation of measurement errors in our models, respectively.
\section{Synthesis imaging with SIM}\label{imsim}
As of November 1998, certain aspects of SIM design which are important
for synthesis imaging have not yet been finalized. The most important
of these is to have a sufficiently complete range of of baseline spacings
available. For the purpose of this
paper, we assume that SIM will have baselines between 0.5 and 12~m in 
increments of 0.5~m, similar to the original design (see
\cite{all97} for an overview). 
With these capabilities, SIM will be well suited to
perform reliable synthesis imaging for a wide range of complicated sources.

The basic ``measurement'' for synthesis imaging consists of determining the 
amplitude and phase of the interference fringes for a given
baseline and spacecraft orientation. These two quantities together form 
the ``complex visibility''. In short, the signal of the two interfering beams
on the detector depends on the path difference or ``delay'' between them. 
The interference pattern is a cosine wave whose amplitude and phase
are the desired observables. If the path delay is accurately known and 
controllable, the complex visibilty can be determined by ``scanning''
the delay. Repeating this measurement for a large number of baselines
then allows a more or less complete reconstruction of the
sky brightness distribution, depending on its complexity. In Appendix A
we describe in more detail the operation of the interferometer in
terms of Fourier optics. Here, we
concentrate on the performance of SIM for astronomical targets.
\subsection{Observing procedure}\label{proc}
\ben
\item{The relative length of the baseline vector is monitored and
controlled with an internal laser metrology system, while its exact 
length and orientation are determined by observing nearby bright stars. 
Pointing stability is achieved with two additional guide interferometers.}
\item{The science interferometer is aligned such that the phase center 
(zero delay, or white light fringe) coincides with the position of a 
reference star.}
\item{The delay lines are adjusted such that the phase center moves to
the science target which has to lie in the 15\deg\ field of regard of SIM. 
In the case of astrometric measurements of
a point source {\it position}, the only interesting observable is the  
delay needed to maximize the white light fringe. This determines the position
of the source with respect to the reference star. Identification of the 
white light fringe is unambiguous since the light is spectrally dispersed: 
only the central fringe will show no phase shift for all spectral channels.}
\item{In the case of an extended source that is to be 
imaged, the complex visibility is measured with 
the delay modulation method (\cite{sha77}) or some variation thereof, which
we describe in Appendix B. Since the signal is spectrally dispersed,
the complex visibilities are measured simultaneously for all spectral channels.
This is the basis for the spectral synthesis technique that we describe
in Section~\ref{specsyn}.}
\item{Baseline length, orientation, and
wavelength determine the \uv -coordinate to which the complex visibility
is assigned. The wavenumbers $u$ and $v$ are 
defined as $u\equiv B_x/\lambda$ and
$v\equiv B_y/\lambda$, where $B_x$ and $B_y$ are the baseline
projections onto the $x$ and $y$ axes of the reference frame.
The measurement is repeated for
as many baselines as necessary to sufficiently fill the \uv -plane.\footnote{ 
The exact meaning of ``sufficiently'' is a non-trivial problem and
depends critically on the (unknown) source structure. This is the reason
why detailed source modeling is needed to predict the imaging performance of
any interferometer.} The image can then
be recovered from the \uv -data by Fourier-inversion, and spurious
responses removed by computer restoration.}
\een
\subsection{Instrumental Parameters}\label{instpars}
\subsubsection{Spatial resolution}\label{resolution}
The spatial resolution of a synthesis imaging observation is not
a priori defined. A diffraction-limited filled aperture 
with a diameter of $D = 12~m$
operating at a wavelength of $\lambda = 500~nm$ has a central response
FWHM of 1.22 $\lambda /D = 0.01$\as . 
In contrast, the final resolution of an image reconstructed from a set of
interferometer measurements is not uniquely defined.
By varying the weighting of the individual \uv -coordinates, or even
removing some baselines completely, one can put more or less emphasis on 
certain spatial frequencies in the reconstruction, and thus change the 
effective resolution, even {\it after} the data are taken.

For example, using a Gaussian weighting which falls to 0.25 at the longest
(12~m) baseline results in a FWHM of 7~mas with very low sidelobes.
Another option is to use only a narrow ring of long baselines 
which yields a resolution almost twice as high as that of the full telescope. 
This fact is regularly exploited in ``aperture masking'' observations 
of bright double stars where collecting area is not an issue, 
but the highest possible resolution is required. Therefore,
SIM will yield a resolution for synthesis imaging 
anywhere between 5 and 10~mas, depending
on the actual distribution of measurements in the \uv -plane, and the
weighting algorithm used.
\subsubsection{Field of View}\label{fov}
The field of view (FOV) is limited by the coherence length of the light, 
$l=c\tau=c/\Delta\nu=\lambda^2/\Delta\lambda$, where $\tau$ is the
coherence time of the light, $c$ is the speed of light, and $\Delta\nu$ is
the bandwidth of the filter used for the observations. 
It can be easily calculated
as the product of the spectral and spatial resolutions, which can
be seen as follows: while the path lengths from the two interferometer elements
are assumed to be equal in the center of the field, an angular separation
$\Theta$ from the center along the baseline vector $\vec{D}$ 
results in a path difference $\Theta\cdot D$. 
As long as this difference is small compared
to $l=\lambda^2/\Delta\lambda$, the fringe contrast will not be affected. 
Thus, $\Theta < \frac{\lambda}{D}\cdot \frac{\lambda}{\Delta \lambda}$, or
$\Theta < R_{\rm spatial\it}\cdot R_{\rm spectral\it}$. 
For example, a 10~nm wide channel 
at $\lambda=600~nm$ would result in a FOV of 0.6\as . Any single
baseline has a FOV that forms a 0.6\as\ wide, long strip on the
sky that is positioned perpendicular to the baseline. However, as the baseline
rotates, sources along the strip more than 0.6\as\ away from the center 
will appear and disappear in the FOV, thus affecting
the measurements in an unpredictable way. In order to avoid this confusion,
SIM will have a circular aperture stop in its optical path.

There is yet another constraint that sets even tighter limits on the
FOV, namely, the need to avoid aliasing. Especially for a complicated
source structure or crowded field
this can be a major limitation on any deconvolution algorithm, since the
sum of the grating rings of all the sources in the field can --- and will ---
mimic fake sources, as will be demonstrated in Section~\ref{clumps}. 
Unless a priori knowledge about the source structure exists, aliasing must 
be avoided for reliable source reconstruction. This means that in practice
the SIM FOV is limited to a circle with a diameter of 0.2\as\ , which
can be seen as follows. The position of the first grating ring
of a point source in the field is determined by the spacing of the baselines.
In the case of 0.5~m increments, the grating ring has
a {\it radius} of 10/0.5=20 times $\lambda /D$ or 0.2\as . To make sure that no 
aliasing occurs from any source inside the FOV, it has to be
limited to a {\it diameter} of 0.2\as . All of our simulations
in this paper, however, have been carried out with a FOV of 0.3\as\ in
order to demonstrate the above effects.

Figure~\ref{fig1} shows a possible \uv -coverage obtained with 170 
measurements and the resulting point spread function (PSF) or ``dirty beam''. 
Here, and in all the simulations that are described in this paper, we
have set the longest baseline to 10~m, which corresponds to the 
requirement of the original SIM design. The current design offers a
maximum baseline of 12~m. 
The grating rings in the PSF described above are clearly visible.
The number of samples in the \uv -plane in Fig.~\ref{fig1} is twice 
the number of 
measurements made. This is because any astronomical object has a purely real 
brightness distribution, so the mutual coherence function of the
incoming wavefront is hermitian. Therefore, 
once the complex visibility of point \uv\ has been measured, its conjugate
can be entered at point ($-u$,$-v$). In addition, different from the case of radio
interferometry, the total flux of the source is determined during every
visibility measurement. This gives the value at the origin of the \uv -plane
(the ``zero spacing'') to high accuracy. 

The ``noisy'' residuals around the central peak of the PSF are due to the
incomplete \uv -coverage. Since, however, the \uv -coordinates of the SIM measurements
are defined with extremely high precision, the shape of the PSF can be
{\it calculated}  very accurately and without the uncertainties and noise
of an actual measurement. It is this precise knowledge of the PSF 
that makes SIM such a powerful imager, since the success of deconvolution 
algorithms like CLEAN or the Maximum Entropy Method (MEM)
is limited by the knowledge of the PSF. 

The accuracy of the ``raw'' synthesized image before deconvolution 
depends in a complicated way on several parameters, including the noise 
properties and flux distribution of the source, 
the stability of the phase measurements, and the actual \uv -coverage. In order
to investigate these issues in detail, we have developed a software toolkit
which we will describe in Section~\ref{simsim}. 
\subsubsection{Sensitivity}\label{sensitivity}
Jumping ahead, we list here the results of the simulations described in
sections \ref{bh} and \ref{exo} in order to provide
an estimate of typical integration times needed to perform synthesis
imaging of extended astronomical objects.

The surface brightness threshold for making
scientifically useful images with SIM in the continuum band from
500-800 nm is $SB_c$ $\approx$ 0.3 mJy/\bs\, which corresponds 
to $m_V$ = 17.7/\bs .
Under the same conditions an emission line such as \ha\ can be imaged
in a narrow (4~nm was assumed in the simulations) band with a threshold 
surface brightness of $SB_l \approx 2.5 \times 10^{-14}$ ergs/cm$^2$/sec/\bs .
A full synthesis with SIM will take typically 4 -- 6 hours of on-target
integration time for sources at these brightness levels. 
During this time, 150 -- 200 different baselines and
baseline orientations will be observed. Observations can of course be
repeated to increase the sensitivity.

These sensitivities can be achieved even in the presence of point sources
up to V=12 in the field of view using the nulling imaging mode 
discussed in the next section.
\subsection{Imaging in nulling mode}
Because of its unprecedented baseline stability, SIM has great potential
for interferometric nulling of point sources. One of the SIM beam combiners
will introduce an achromatic 180\deg\ phase shift in one of the 
inteferometer arms by polarization inversion, thus eliminating the light 
from a point source that is located at the phase center. 
This capability is a programmatic technology demonstration requirement.
The experience gained is crucial for the design of future space missions 
that are aimed at the detection of extra-solar planets, like e.g. the 
Terrestrial Planet Finder (TPF). The requirement for SIM is to reach a 
nulling efficiency of $10^{-4}$. Many interesting science programs intend 
to use this capability to determine the spatial extent of objects like
stars or supernova shells by measuring the light leakage around the phase
center.

In this paper, we will describe another mode of observation, namely
synthesis imaging in nulling mode. In Section 5, we will discuss the specific
features of this technique for the case of exo-zodiacal dust disks.
\subsection{Spectral synthesis imaging}\label{specsyn}
Since SIM is designed to perform astrometric measurements in a dispersed fringe
mode, it has a built-in spectroscopic capability. That is to say, the
amplitude and phase of an object's wavefront are monitored simultaneously
at a number of different wavelengths. Since the same physical baseline 
translates into different \uv -coordinates at different wavelengths,
this information can, in principle, be used to increase the \uv -coverage.
This method is usually referred to as spectral synthesis imaging.

Spectral synthesis has one major caveat: the implicit assumption in the
reconstruction of the source is that its
{\it structure does not change with wavelength}. In the optical regime
this assumption is far from safe. In fact, it is often plain wrong, 
especially when strong emission lines are present in the bandpass considered. 
That does not mean that one cannot use spectral synthesis with SIM at all.
There are classes of objects for which the above assumption is
acceptable. For example, in the case of zodiacal disks around 
solar-type stars, it may be reasonable to assume that the structure 
of the source does not vary strongly over the SIM passband, 
since it is observed only in reflected 
starlight. We will present the results of spectral synthesis simulations 
for exo-zodiacal disks in Section~\ref{exo}.
\section{{\sc SIMsim} --- a software toolkit}\label{simsim}
In order to simulate the imaging capabilities of SIM, we have
developed \ss , a software package written in the Interactive Data Language 
(IDL) environment. The motivation was to obtain realistic performance
estimates for a variety of target sources,
and to develop optimal observing strategies. The principal steps carried
out during a simulation are illustrated in Fig.~\ref{fig2}.

In brief, a model source distribution is put on a finely-sampled grid (typically
$256\times 256$ pixel). For simplicity, the field center is assumed 
to be the center of rotation of the spacecraft during the synthesis imaging,
and no effort has been made so far to account for possible pointing instabilities.
The mutual coherence function of the source wavefront is calculated via a
Fast Fourier Transform (FFT), and sampled on the \uv -coordinates observed
by SIM. Noise is added, and the actual phase and
amplitude retrieval process is simulated. The details of this algorithm are described
in Appendix B. So far, we have incorporated photon noise
according to Poisson statistics, and random phase calibration errors with
an amplitude of up to 0.5\deg . A final Fourier inversion of this
sampled and corrupted data then gives the 
``dirty'' image. The PSF is calculated by transforming a mutual coherence
function of unit amplitude and zero phase at each \uv -sample point
of the observation. Deconvolution algorithms like CLEAN or the Maximum 
Entropy Method (MEM) 
can then be applied to minimize the effects of the incomplete \uv -coverage.

The nulling capability of SIM is modeled as follows. Before the 
noise is added to the source wavefront, its amplitude $A$ (at each \uv -coordinate)
is reduced by most of the amplitude $A_c$ of the source at the phase center
$u=v=0$:
\beq \label{null_eq} 
A_{nulled} = A - A_c\cdot(1-\epsilon)
\eeq
where $\epsilon$ denotes the accuracy to which nulling can be achieved. 
For SIM, $\epsilon$ is expected to be around $10^{-4}$. It is
limited by stray light due to imperfections in the optical elements and
by the stability of the pathlengths. The phase of the wavefront is set to zero,
resulting in an intrinsically symmetric source reconstruction. The reason
for the loss of phase information is that the delay cannot be modulated
without losing the null: once the light is minimized, the delay lines have
to be kept as stable as possible. The reduced amplitude of the signal
results in lower photon counts on the detector which, in turn, reduces
the poisson noise.

Equation \ref{null_eq} ensures that the total flux in the reconstructed
image is equal to the flux in the whole field-of-view {\bf except for
the central pixel} which, in the simulations, contains the flux of the
nulled star. This is, of course, known for the models. However, in a
real observation, such {\it a priori} information on the exact value of
$A_c$ does not exist.  It is only possible to measure the total flux --
including the central source -- by doing a standard observation in the
direct beam combiner. An iterative approach will be necessary, which
adds an appropriate {\bf constant} to each amplitude measurement; this
constant is an estimate of the total flux in the field, minus the flux
from the central source. If this value is chosen correctly, the
reconstructed image will be zero at the phase center.

Since the model sources are intrinsically constructed on a regular grid, as
are the \uv -coordinates of the ``measurements'', there is no need for 
gridding procedures in our code. However, real sources and real observations
will not be so conveniently organized, so that in order to fully model the 
SIM observations, weighting and gridding algorithms will have to be 
implemented. This work is currently underway.

Figure~\ref{fig3} illustrates the Graphical User Interface (GUI) of
\ss . The top panel allows the user to select between a number of model sources and
to choose various observational parameters of SIM such as \uv -coverage, 
spectral resolution, etc.
The intermediate steps of the simulation are displayed in the four graphics
windows. All results can be saved in postscript format for printing or in
FITS format for further work with IDL or any other data analysis environment.
In the following sections we use \ss\ to model the capabilities of SIM
to image two interesting classes of astronomical objects.
\section{Black holes in galactic nuclei}\label{bh}
Black holes have been detected convincingly in at least a dozen galaxies,
both active and quiescent. The techniques for such detections are multifold
(see \cite{kor95} for a review), but have concentrated on
the dynamics of the gas and stars in the neighborhood of the black hole.

Stellar kinematical studies are often hampered by uncertainties
regarding the stellar velocity dispersion anisotropy, and the need for
high spatial resolution in connection with intrinsically low surface
brightness makes such measurements very challenging even with HST. 
However, some galaxies have circumnuclear disks of ionized gas
which provide a more favorable tool for black hole detection.  The
emission lines from the ionized gas are bright and have large
equivalent width, and the dynamical state of the gas is relatively
simple; its rotation curve directly yields the black hole mass.

As a starting point for our discussion, we have chosen M~87 as the best
known example for such a line-emitting gaseous disk. The next section
briefly describes our input model for M~87.
\subsection{An example: M~87}\label{m87}
M~87, a dominant giant elliptical galaxy in the Virgo cluster,
is the prototype for galaxies that harbour a central black hole, the presence
of which is revealed by a line-emitting gaseous disk. The nucleus of 
M~87 has been imaged with HST in both continuum (\cite{lau92}) and line 
emission (\cite{for94}, and the ionized gas has been studied
spectroscopically in great detail (\cite{har94,mac97,mar97})
We thus have a fairly clear picture of the properties of the nuclear gas disk. 
Table \ref{tab1} summarizes the parameters we used to construct a realistic model 
flux distribution for the gaseous disk around the central black hole in M~87.
The \ha\ surface brightness over the central 0.25\as\ of our model is 
$1\cdot 10^{-12}$~ergs/cm$^2$/\AA\/\bs , as derived from the observed 
[SII] flux (\cite{har94}) and an assumed line ratio of [SII]/(\ha +[NII]) = 4.
The gas is assumed to show Keplerian rotation according to
\beq \label{rotvel_eq}
(\frac{v_c}{100~km/s})^2\,=\,0.08869 \cdot \frac{Mpc}{d} \cdot 
	\frac{M_{\bullet}}{10^6 M_{\odot}} \cdot \frac{arcsec}{r}
\eeq
where \mbh\ denotes the black hole mass, $r$ the angular distance from the center,
and $d$ the distance to the galaxy. The numbers listed in Table \ref{tab1} then 
result in the rotation curve shown in Fig.~\ref{fig4}.

The \ha\ emission from the rotating disk would appear in a range
of spectral channels, according to the Doppler shift of the gas velocity
projected along the line of sight. For the above parameters, a (hypothetical)
diffraction-free 10m-telescope in space, equipped with an integral field 
spectrometer with 4~nm wide channels, would see the source as in Fig.~\ref{fig5}. 
Here, we have convolved the ``true'' flux distribution of the M~87 model with 
the ``clean beam'', a Gaussian fitted to the central part of the PSF.
While the central, unresolved continuum source appears in all channels,
the appearance of the \ha\ emission depends on parameters like
channel width, black hole mass, inclination of the disk, etc.

To demonstrate the power of our analysis tools, we present in the next section
what can be expected from SIM observations of M~87 and similar sources.
\subsection{Results of the {\sc SIMsim}ulations}\label{results}
In Fig.~\ref{fig6}, we present a few examples of our simulations that are intended 
to give an impression of what imaging with SIM can do in this context.

The first row of the left panel again shows those channels of our M~87 model
containing \ha\ emission. The channels again are 4~nm wide and centered at
554, 558, and 662~nm, respectively\footnote{We have masked the central 
continuum source in these plots for better display only; the observation
was not simulated in nulling mode.}.
The second row shows the result of the direct Fourier inversion of the
complex visibilities measured by SIM. The \uv -coverage was the same 
as in Fig.~1, and the integration time was 100s per measurement, for a total
on-source integration time of about 5~hours. The nominal SIM throughput of 0.3 was
assumed. We added photon noise to the
measurements according to Poisson statistics. The third row shows the 
results after 200 iterations with the MEM algorithm. The
emission structure of the high velocity gas is well recovered.
The same is true for the \ha\ intensity profile, as shown in Fig.~\ref{fig7}. 

The spatial extent of the \ha\ emission in each channel provides valuable
information on the velocity gradient, and thus the black hole mass.
This is demonstrated in Fig.~\ref{fig8}, where we have
run the same simulation as described above, but for various black hole masses
\mbh . We show only the blueshifted wing of the \ha\ emission for values of
\mbh\ = 2, 3, and 4$\times 10^9$~\msol . The SIM observations are very 
sensitive to the value of \mbh , a direct consequence of the linear relationship
between black hole mass and radial distance in a Keplerian rotation
curve, cf.~eq. \ref{rotvel_eq}. However, it should be stressed that 
in practice fitting
procedures are required to solve for all the free model parameters including
black hole mass, disc inclination, and the actual distribution of
disk surface brightness. 

We emphasize that no attempt has been made to optimize the deconvolution 
process. For example, the stability and convergence of MEM can be greatly
improved by using ``a priori'' knowledge about the source structure. However,
in order to better be able to compare the results for different source models, 
we have --- for all the simulations in this paper --- started the MEM deconvolution 
with an empty array, and stopped after 200 iterations. For the real SIM
observations, one would make every effort to optimize the deconvolution 
strategy for each individual dataset. This is, in fact, another area where future
simulation work with \ss\ will be useful. However, it is our experience
that if no trace of the source structure is evident in the direct Fourier 
inversion, no improvement can be expected from deconvolution, regardless of
the algorithm employed. Thus, the quality of the data can be most directly 
judged from the direct Fourier inversion itself, before any restaoration.
\subsection{Is there more than M~87?}\label{more}
An important goal of \ss\ is to define requirements on various aspects of
the SIM spacecraft. As an example, what is the minimum spectral resolution 
of the SIM beam combiner that allows observations such as the one discussed
here? Certainly, the channel width must not be smaller than the
wavelength shift of the emission line under observation at a distance
of at least one resolution element from the center. This, of course, 
depends on the black hole mass and the distance to the host galaxy. 
In Fig.~\ref{fig9}, we have 
plotted the limiting black hole mass (i.e. the minimum mass that causes 
a shift of the \ha\ emission of at least one spectral resolution
element at an angular separation of two spatial resolution elements
(0.02\as ) from the center) as a function of the distance to the host
galaxy for a number of possible SIM spectral resolutions and
disk inclinations. For example, with 1~nm wide spectral channels (dotted lines),
SIM can potentially detect $10^8$\msol\ black holes up to a distance of 
$\approx$ 80~Mpc, if the disk
has an inclination $i \leq 20^\circ$. Whether the \ha -flux in cases
different from M~87 is sufficient 
to acquire a high signal-to-noise ratio in a reasonable amount of time
is more uncertain. 
\section{Exo-zodiacal disks} \label{exo}
Circumstellar dust disks are commonly thought to provide the material
for planet formation. The structure of such disks is therefore important
for our understanding
of the processes that lead to the formation of planetary systems.
Many searches have been started
for such disks, most of which are based on coronographic observations and
concentrate on stars with a high
IR-excess as measured by the IRAS satellite (e.g. \cite{smi92}). 
Until most recently, however, only one unambiguous detection 
of a circumstellar dust disk outside the solar system had been reported, 
namely the \bpic\ system. From modeling the \bpic\ disk,
\cite{kal96} found that it is the uncommonly large scattering cross-section 
of the dust particles that makes \bpic\ a special case. If there were other
nearby systems like \bpic, they would have been detected regardless of the disk 
inclination. \cite{kal96} conclude that at least an order of magnitude 
improvement in the suppression of the stellar light is needed to detect 
more ``normal'' circumstellar disks, i.e. with scattering cross section 
less than a tenth of that in \bpic .

SIM, with its interferometric nulling capability, differs from groundbased
coronagraphic observations in a number of ways: 
\ben
\item{The starlight suppression in SIM is not limited by atmospheric seeing or the
efficiency of an adaptive optics system, but only by the quality of the
optical elements and the pathlength stability.} 
\item{The high spatial resolution of SIM allows the 
detection of disks much smaller than that in \bpic , and SIM can therefore 
image such systems
to greater distances $d$. In fact, SIM's ability to detect exozodiacal
disks increases with $d$, as long as the disk fills 
the 0.2\as\ FOV of SIM. This somewhat surprising
behavior is due to the fact that the surface brightness of
the disk varies only slightly with $d$, so its total signal on 
the SIM detector will stay more or less constant. 
The signal from the star, on the other hand, 
decreases as $1/d^2$, so that the noise contribution from the
nulling residuals of $10^{-4}$ become less
important. In addition, the further away the star is, the more it resembles
a true point source. This will reduce light leakage around the null, which is
also a source of photon noise even for a perfect nulling beam combiner.}
\item{Since all phase information in the nulling process is lost, the
source after reconstruction {\it always} appears symmetric. That means that
eventual asymmetries in the disk can not be recovered.}
\item{Aliasing due to the incomplete sampling of the mutual coherence
function can sometimes mimic companion stars or planets.}
\een
In what follows, we investigate the SIM potential for the detection of 
zodiacal disks and specifically address the points listed above.
\subsection{The model} \label{model}
We use a simple model of a circumstellar dust disk, based on the
zodiacal emission of the solar system (\cite{rea96,tra98}). 
The particle number density of the solar system zodiacal dust cloud can
be well modeled by the expression
\beq
\rho = \rho_0 \cdot r^{-\alpha} \cdot e^{-\beta \cdot (z/r)^{\gamma}}
\eeq
where $r$ is the in-plane radial distance from the center, $z$ is the 
modulus of the perpendicular-to-plane distance (in AU), $\alpha = 1.4$, 
$\beta = 3.26$, and $\gamma = 1.02$. We treat the scattering efficiency of
the dust particles as a flux scaling parameter, and
we assume isotropic scattering. In what follows, we investigate 
the two limiting cases of
a face-on ($i$=90\deg ) and edge-on ($i$=0\deg ) orientation of the disk.

The radial surface brightness distribution of the face-on disk can be
derived from integrating the volume density. For $\gamma=1$ it is
given by the expression
\beq
\sigma_{(i=90^{\circ})} = C\cdot \sigma _1 \cdot r^{-\alpha + 1} .
\eeq
Here, $C$ is a scaling factor that characterizes the scattering 
efficiency of the dust particles in an exo-solar system relative to the
solar system.
The normalization constant $\sigma_1$ is the surface brightness of the 
face-on disk, as seen from outside the solar system, 
at a distance of 1~AU from the sun. Assuming isotropic scattering, this
number is (at $\lambda$=550~nm)
\beq
\sigma_1 = 0.13~\rm MJy/sr .
\eeq
An approximate ratio of the surface densities of the edge-on and face-on
directions at a given value for $r$ can be obtained by ratioing the 
characteristic scale heights
of the $z$-direction and along a tangent to $r$ in the plane:
\beq
e/f = \beta^{1/\gamma}\cdot \sqrt{e^{2/\alpha}-1} \approx 5 .
\eeq
An approximation to the edge-on surface brightness distribution 
is therefore
\beq \label{eq1}
\sigma_{(i=0^{\circ})} = C\cdot e/f\cdot \sigma_1 \cdot r^{-\alpha + 1}\cdot e^{\frac{-\beta z}{r}} .
\eeq
where $\gamma \approx 1$ is used. Figure~\ref{fig10} gives a visual
impression of the model in both the face-on and edge-on case.
The spectrum of the
dust in the wavelength range discussed here (500 -- 1000~nm) is assumed to
be that of a black body with the temperature of the central star.
\subsection{\bpic\ --- the showcase} \label{bpic}
In this section, we investigate the question whether SIM is capable of 
finding other systems like \bpic , should they exist. In this context,
we define ``\bpic -like'' as having about 1000 times the dust content
and surface brightness of the solar system, i.e. $C$=1000, without 
any detailed modeling of the specific properties of \bpic\ itself. 

As mentioned above, the SIM performance in terms of imaging of 
circumstellar disks is very limited for nearby systems, but improves with 
increasing distance, as long as the disk fills the FOV. 
For example, the \bpic\ disk itself is impossible to image with SIM, 
because even with $10^4$ nulling, the photon noise of the remaining 
light from \bpic\ would make the detection of the faint disk emission 
impossible. Figure~\ref{fig11} demonstrates the increasing sensitivity of SIM 
up to distances of 1~kpc. 

Again, we note that no attempt has been made to optimize the restoration
procedure in order to avoid additional degrees of freedom. As can be seen in
Figure~\ref{fig11}, SIM can easily detect \bpic -like systems up to a distance
of about 1~kpc, beyond which our model disk no longer fills the SIM FOV and 
its total flux drops rapidly. This distance limit depends linearly on the 
extent of the
disk, for which we have assumed a value of 100~AU. This is a somewhat 
arbitrary number, chosen such 
that the angular diameter at a distance of 1~kpc is 0.2\as, equal to the SIM 
FOV. The central star is assumed to be of solar type ($M_V$=5).
\subsection{Can SIM find other solar systems?} \label{solar}
The above results change dramatically if systems with lower dust contents
than \bpic\ are considered. The dust content of our model disks is parametrized
by the constant $C$ in eq. \ref{eq1}. Figure~\ref{fig12} summarizes our results on
studying various values for $C$. Disks with more than a tenth of
the dust content of \bpic\ can be detected with SIM. However, it is evident 
that zodiacal disks similar to that of the solar system will not be found 
with SIM, regardless of their distance or inclination. 
\subsection{What are the clumps?} \label{clumps}
It is interesting that in all of our reconstructions, the disk structure after
deconvolution appears much more clumpy than the smooth input model. In fact,
when looking at the results, one might in many cases think to have found a
number of planets. These ``clumps'' are artefacts of the restoration
process on our incomplete
\uv -data. While the sidelobe pattern itself is 
known to high accuracy, the poisson noise associated with the signal does
not have the shape of the PSF. 
Spikes in the noise distribution create residuals 
that are mistaken as real source structure by any
deconvolution algorithm. The fact that the clumps are indeed randomly
distributed can be seen when comparing three independent simulations of
the face-on zodiacal disk of a \bpic -like system at 1~kpc and a nulling
efficiency of $\epsilon = 10^{-4}$ in Figures~\ref{fig11},\ref{fig12}, and
\ref{fig13}, which only differ in their random photon noise. A repetition
of the observation would allow identification of such spurious objects.
\subsection{How critical is the Nulling?} \label{null}
In order to investigate the effects of a degradation of the SIM
nulling capability, we have run a series of simulations with different values
for $\epsilon$, the nulling efficiency in eq. \ref{null_eq}. Figure~\ref{fig13}
demonstrates that for a \bpic -like system at 1~kpc, the most favorable case
in all of the simulations we have run so far, SIM could detect the disk even if
it had only $\epsilon=10^{-3}$. However, anything worse than that would not
suppress the starlight sufficiently even at such high distances. We conclude that
in order to be a powerful search engine for exo-zodiacal disks, SIM needs to achieve
at least $\epsilon=10^{-4}$, and anything better would improve the performance
considerably.
\subsection{Planetary Gaps} \label{gaps}
A massive planet --- if present --- is believed to sweep up any 
inward-spiraling debris and thus cause a dust-free region in the 
circumstellar disk inward of the planet's orbit. Such an ``inner gap'' 
is believed to exist in the \bpic -system, as inferred from its spectral 
energy distribution (\cite{lag94}), but the gap has never been directly 
imaged owing to the large radius which is 
masked out by coronographic instruments.

With its high spatial resolution and sensitive nulling, SIM is able
to directly image possible inner clear regions in exo-zodiacal systems.
This is demonstrated in Fig.\ref{fig14} where we have investigated
the case of a \bpic -like disk at a distance of 500~pc with and 
without a 20~AU inner clear region. Clearly, the signature of
the disk depends on its size and the distance to the system, but
a 20~AU radius gap would be detectable in systems as distant as 1~kpc. 
Here, we just want to illustrate the intriguing prospect of directly
imaging such gaps (and later, possibly the planets that cause them).
\section{Summary} \label{sum}
We have built an interactive software package that enables us to
perform numerical simulations of imaging observations with SIM.
The results obtained so far can be summarized as follows:
\ben
\item{SIM, with a spectral resolution of 1~nm, can resolve the velocity structure of
line-emitting gaseous disks around black holes in the centers of nearby galaxies down
to black hole masses of $10^8$\msol\ at distances up to 100~Mpc. With this
resolution, SIM is in principle capable of determining the 
mass of the black hole to 10\% accuracy. The necessary integration time depends
on the \ha -flux from the disk; for cases like M~87, about 5~hours is sufficient. 
A coarser spectral resolution of SIM would constrain the accessible
parameter space for quantities such as distance, black hole mass, or 
disk inclination.}
\item{Given a nulling efficiency of $10^{-4}$, SIM can image zodiacal disks
with dust contents similar to that of the \bpic -system with high 
signal-to-noise ratio in about
5~hours of integration time. The limiting distance depends only on the angular 
size of the disk, which should be no smaller than 0.1\as .}
\item{Inner clear regions in the disk can be directly detected and
imaged with SIM for systems at a distance of up to 1~kpc provided that the 
dust-free region has a radius of at least 20~AU.} 
\een

We are grateful to R. van der Marel
for his advice on the black hole simulations, and to W. Traub for
his expertise on the zodiacal disk model. We also would like to thank
S. Unwin, M. Colavita and M. Shao for their interest in and continued 
support of this project. The work described in this paper
is carried out at STScI with financial support from the SIM project at 
the Jet Propulsion Laboratory (JPL). 
\begin{appendix}
\section{The Fourier optics of SIM}
The idea that an interferometer could be constructed to measure the
angular diameters of stars is usually credited to \cite{fiz68},
although no details were provided in that particular reference. Twenty
years later, \cite{mic90} took it up and,
another 30 years later with Pease, constructed the now-famous stellar
interferometer on the 100-inch telescope (\cite{mic21}).
Figure \ref{fig15} shows a sketch of the original drawing from that
paper, to which we have added symbols indicating the spacing $L$ of the
outrigger mirrors and the spacing $D$ of the mirrors at the telescope
aperture which direct the light to the primary mirror. Variations of
this sketch appear in many introductory textbooks on physics and
astronomy.

The image of a star in the focal plane of the stellar interferometer
is crossed by interference fringes, as shown in cross section in 
Fig.\ref{fig16}, where some additional parameters are defined: the {\it
fringe amplitude} $A$, the {\it fringe period} $\Delta \theta$ and the
{\it fringe phase} $\phi$. In this picture the average signal from the
star has been normalized to 1. The overall size of the Airy pattern is set by 
the aperture formed by the cascade of small reflecting
mirrors which direct the light to the telescope primary mirror.

As is well known from the general theory of wave diffraction, this
Fraunhofer diffraction pattern in the focal plane of the telescope is
the square modulus of the Fourier Transform of the distribution of
illumination (in amplitude and in relative phase) over the aperture.
The expression for the stellar interferometer can be derived in a
straightforward way, at least in one dimension, by starting from the
result for the single-slit diffraction problem. We then use standard
theorems from the theory of Fourier Transforms, in particular the shift
theorem and the convolution theorem, and apply them to a plane wavefront over
the entrance aperture (pupil). In this way it is straightforward to
show that the final Fraunhofer pattern for an unresolved star is the
square modulus of
\begin{eqnarray}
 M(u) & = & F(u-u_0) \times G(u-u_1) \\
     & = & 2\,\sinc(\pi a(u-u_0))\cos(\pi (Du-Lu_0)) \nonumber
\end{eqnarray}
with $\sinc(x) \equiv \sin(x)/x$, and $u \equiv sin\phi$, $\phi$ being
an angle as defined in Fig.\ref{fig16}.
This is the product of the pattern $F(u) = \sinc(\pi a(u-u_0))$ 
for a single slit of
width $a$ offset to $u_0 = \sin\phi_0/\lambda$, and the pattern 
$G(u) = 2 \cos(\pi D (u-Lu_0/D))$
for a very thin double slit offset to $u_1 = Lu_0/D$, where $\phi_0$ is the
angular distance of the star from the direction normal to the baseline $L$.
We note that:
\begin{itemize}
\item As the source position $\phi_0$ changes, the fringe phase
moves at an amplified rate $Lu_0/D$. The positional accuracy is therefore
{\it magnified} by $L/D$.
\item The fringe period remains set by the separation $D$ of the
secondary apertures no matter what is the separation $L$ of the primary
apertures on the outriggers.
\end{itemize}
SIM can be regarded as a Michelson stellar
interferometer with two specific features:
\begin{itemize}
\item{The separation $D$ of the secondary apertures is reduced to
zero\footnote{Practical methods of achieving this beam combination usually
introduce an additional fringe phase of typically $\approx \pi/4$ which
we ignore here.}. The fringe period therefore becomes infinitely
large, and the entire image of the star is either light or dark
depending on the fringe phase.}
\item{A delay $\Delta = L\sin\phi_0 + \delta$ is added to one side
of the interferometer light path (e.g. between M$_1$ and M$_2$ in order
to shift the double-slit pattern from $u_1$ back to within a small 
offset $\delta$ from the position of the
star $u_0$. This is necessary because of the finite bandwidth of the
signal, which results in a limited coherence length, as discussed in 
Section \ref{fov}. We will discuss the bandwidth effects on the fringe
contrast later in this section.}
\end{itemize}
The Fraunhofer pattern for a monochromatic point source in our 
one-dimensional model of SIM is therefore the square modulus of:
\beq \label{eqA2}
M(u) = 2 \, F(u-u_0) \, \cos(\pi (L\sin \phi_0 - \Delta)/\lambda),
\eeq
which is
\beq \label{pointsource}
P(u) = |M(u)|^2 = P_0(u)\,(1 + \cos(2\pi \delta/\lambda))
\eeq
where $P_0(u) \equiv 2\,|F(u-u_0)|^2$.
Viewed as a function of the delay offset $\delta$, this is a
co-sinusoidal pattern with an average value set by the intensity $P_0$
of the star\footnote{The residual factor of 2 present in 
equation \ref{eqA2} is ignored further here. It comes
in because we have doubled the collecting area by using two outrigger
mirrors, and it will be subsumed into $P_0$ by the overall
interferometer calibration procedure.}, and varying from zero to
$2\,P_0$. The fringe amplitude here is 1, appropriate for a
point source.
 
Consider now an extended source in the sky, which we
integrate in a direction perpendicular to the interferometer baseline
in order to produce a 1-D image $B(\phi)$ where $\phi$ is an angle
measured in the plane of the baseline (Fig. \ref{fig17}).
Let us further assume that the position $\phi_0$ to which we have 
initially offset the fringe pattern from our calibration grid 
star (by adding a delay $\Delta$) is somewhere in the middle of 
$B(\phi)$. Each part $d\phi$ of the
spatially-extended source then contributes a fringe pattern:
\beq \label{fringecontrib}
dP(\phi) = B(\phi)d\phi\,(1 + \Re e^{-2\pi i(L\sin \phi - \Delta)/\lambda)});
\eeq
where again $\Delta = L \sin \phi_0 + \delta$ and 
$\Re$ denotes the real part of the exponential. Let us further simplify
the notation by assuming that the FOV is confined to a small
region on the sky, so that $\sin \phi \approx \phi$, and measure $L$ in
wavelengths, $l = L/\lambda$. The total fringe pattern obtained by
integrating equation \ref{fringecontrib} is then:
\beq
P = B_0 \Delta \phi + 
\Re\int_{-\infty}^{+\infty}B(\phi)e^{-2\pi i
(l\phi-l\phi_0-\delta/\lambda)}d\phi .
\eeq
We can write the integral in this equation as:
\beq
B_0 \Delta\phi \times \Re e^{2\pi i(l\phi_0+\delta/\lambda)}
\times \Gamma(l),
\eeq
where we recognize $\Gamma(l)$ as the Fourier transform of the normalized $B(\phi)$:
\beq
\Gamma(l) = \frac{\int_{-\infty}^{+\infty}B(\phi)e^{-2\pi i l \phi}d\phi}{B_0 \Delta\phi}.
\eeq
Since $B(\phi)$ will in general not be symmetric about $\phi_0$,
$\Gamma(l)$ will be complex, so we can write
$\Gamma(l) = A(l) e^{2\pi i \phi(l)}$, leading to:
\beq \label{spacefringe}
P = P_0 \{ 1 + A(l) \cos(2\pi(l\phi_0 - \phi(l) + \delta/\lambda))\}.
\eeq
There are 3 unknowns here: the average power $P_0$, the fringe amplitude
$A(l)$, and the fringe phase $\phi(l)$ (measured here in turns).
They can all be determined by measuring $P$ as a function of delay $\delta$.
In practice, this can be achieved by continously
scanning $\delta$ over $\pm \lambda$, and recording $P$ over a number of
time intervals. This is the basis for our simulation of the measurement process
and the introduction of photon noise, as will be described in appendix B. 

Let us now consider the effects of the finite bandwidth of the interfering light.
In order to keep the algebra manageable, we consider only the case of
a plane wave from a point source.
The fringe pattern we have derived in equation \ref{pointsource} 
will be different for different values of the wavelength $\lambda$. 
The total response is the sum of the fringe
patterns for all wavelengths over the spectral range of the passband.
Suppose the passband has a spectral shape $S(\nu)$.
Then, in analogy to equation \ref{fringecontrib}, that part of the spectrum at
frequency $\nu$ contributes a fringe:
\beq \label{spectrafringe}
dP(\nu) = S(\nu)d\nu (1 + \Re e^{2\pi i(L\sin\phi_0 - \Delta)/\lambda}).
\eeq
where $P_0 = S(\nu)d\nu$. We define a delay time $\tau = \delta/c$ where $c$ is the
speed of light. Then we have:
\beq
dP(\nu) = S(\nu)d\nu \{1 + \Re e^{-2\pi i\nu\tau}\}.
\eeq
The total fringe pattern is obtained by integrating:
\beq
P = S_0\Delta\nu + \Re \int_{-\infty}^{+\infty}S(\nu)e^{-2\pi i\nu\tau}d\nu.
\eeq
We recognize the integral as the Fourier transform $s(\tau)$
of the filter function $S(\nu)$. As an example, let's take a rectangular
filter of height $S_0$ and width $\Delta\nu$ centered at $\nu_0$. The
transform of this function is related by the shift theorem to the
transform of a function $C(\nu)$ having the same shape but centered at
the origin. The transform of $C(\nu)$ is
$c(\tau)=S_0 \, \Delta \nu \, \sinc(\pi \tau \Delta \nu)$,
so the transform of
$S(\nu) = C(\nu - \nu_0)$ is $s(\tau) = e^{2 \pi i \nu_0 \tau} c(\tau)$, and the
fringe pattern becomes:
\beq
P(\nu) = S_0 \Delta \nu \{1 + \sinc(\pi \tau \Delta \nu)
\cos(2 \pi \nu_0 \tau) \},
\eeq
so that if we scan $\delta = c\tau$ and record the intensity of
the stellar image in the focal plane of SIM we will obtain the pattern shown
in Figure \ref{fig18}.
\section{Simulation of the measurement and its noise}
In order to provide an estimate of SIM sensitivity and
image quality, various noise components associated with the
measurements need to be modeled as realistically as possible.
The systematic (calibration) errors of the spacecraft are hard to
predict at this time, because the design has not been finalized and a
detailed hardware model is not yet available. As of now, we can only
make assumptions on parameters such as fringe phase stability. 
On the other hand, the statistical errors due to photon statistics
are straightforward to model. They are implemented in \ss\ as follows.

The fundamental shape of the signal as a function of path difference
(delay) between the two interfering beams has been derived in equation
\ref{spacefringe}. Here, the term ``signal'' means the total
number of photons falling on the detector over the integration time.
If the delay $\delta$ is continously modulated over $\pm \lambda$, then
one can integrate the detector signal over four time ``bins'', during
each of which the delay moves over a quarter wave. This method has first been
described by \cite{wya75} and was suggested for use in astronomical
interferometers by \cite{sha77}. It is usually
referred to as the ``quadrature method'' and has been used in many ground-based
interferometers. The usual implementation uses a triangular
modulation of the delay with piezo-electric elements (e.g. \cite{sha88}). 

In ground-based interferometers the delay is scanned over the range of
one wavelength in about 10~ms owing to the short coherence time of the 
atmospheric seeing, which otherwise moves the fringe phase $\phi$ in a random 
way. In principle, there is no need for high modulation frequencies in
space, so that instead of a continous modulation, one could envisage a
``stop-and-stare'' mode for the delay $\delta$. However, since the
continous modulation technique is well established and 
simple to handle mathematically, we have used it to simulate the measurements
in \ss . Following \cite{wya75}, the signal in the four bins 
$B1$...$B4$ which, for mathematical simplicity have the limits
$[-\lambda /8 ,\lambda /8]$,$[\lambda /8 ,3\,\lambda /8]$,
$[3\,\lambda /8 ,5\,\lambda /8]$, and $[5\,\lambda /8 ,7\,\lambda /8]$
can calculated by integrating equation \ref{spacefringe} over these
intervals, yielding:
\begin{eqnarray*}
B1_{u,v} & = & P_0/4 + 1/(\pi\,\sqrt{2})\,A(u,v)\,\sin \phi(u,v) ; \\
B2_{u,v} & = & P_0/4 + 1/(\pi\,\sqrt{2})\,A(u,v)\,\cos \phi(u,v) ; \\
B3_{u,v} & = & P_0/4 - 1/(\pi\,\sqrt{2})\,A(u,v)\,\sin \phi(u,v) ; {\rm and}\\
B4_{u,v} & = & P_0/4 - 1/(\pi\,\sqrt{2})\,A(u,v)\,\cos \phi(u,v).
\end{eqnarray*}
$P_0$ here is the total signal (in photons) in the FOV 
over the spectral passband during the integration time. In the case
of a simulation, the values for the fringe amplitude and phase,
$A(u,v)$ and $\phi(u,v)$, are known from the Fourier transform of the
model source structure. Therefore, the values $B1-B4$ can be calculated,
and poisson noise can be added to them, before using the ``measurements''
to recover the source. 

The noisy bin signals are then used to determine the ``measured''
fringe amplitude and phase according to
\begin{eqnarray*}
A_{\rm noisy\it} & = & \pi/\sqrt{2} \times\sqrt{(B1-B3)^2 + (B2-B4)^2} \\
\phi_{\rm noisy\it} & = & {\rm atan \it}(\frac{B1-B3}{B2-B4})
\end{eqnarray*}

These values are re-entered in an empty array (the \uv -plane), 
and the (sparsely-filled) array is Fourier back-transformed to yield the
reconstructed image of the source.
As an option for the user, an additional noise component can be added that
tries to model the expected phase calibration errors. Our simulations have shown, 
however, that up to the specified maximum error of 0.5\deg , this has negligible
effects on the restoration quality. Since the PSF can also be computed, 
the recovered image can further be
improved by applying restoration algorithms like CLEAN or MEM.
\end{appendix}


\clearpage
\figcaption[fig1.ps]{Left: example of a specific
\uv -coverage, obtained from 170 SIM 
measurements. Right: the resulting point-spread-function. \label{fig1}}
\figcaption[fig2.ps]{Schematic view of \ss\ capabilities . \label{fig2}}
\figcaption[fig3.ps]{Graphical User Interface of \ss . \label{fig3}}
\figcaption[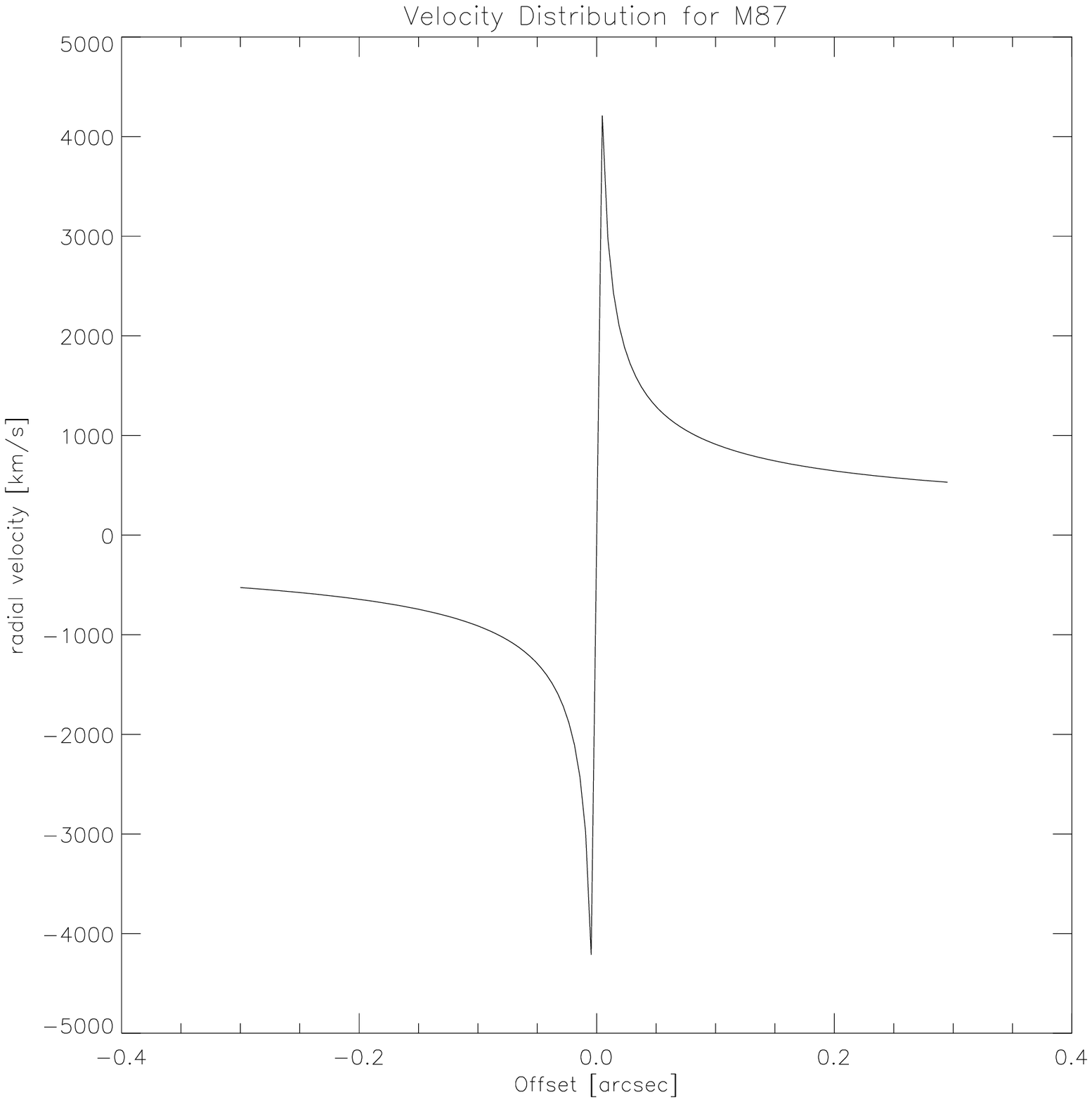]{Model rotation curve for M~87. \label{fig4}}
\figcaption[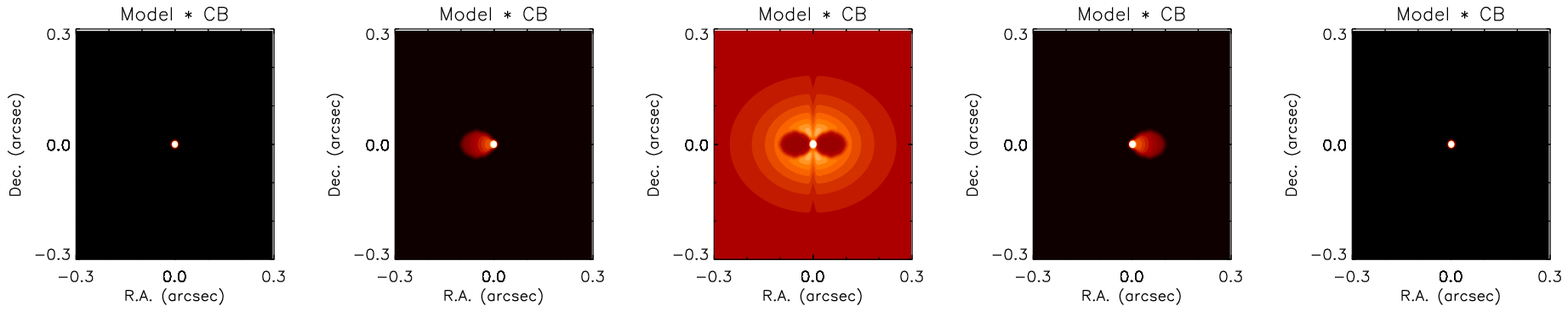]{Model surface brightness distribution after convolution 
with the ``clean beam'' for the \ha\ line emission from a rotating gas disk
around the black hole in M~87. The channels are 4~nm wide and centered at
[-8,-4,0,4,8]~nm offset from the line center. \label{fig5}}
\figcaption[fig6.ps]{Results of the simulated SIM synthesis imaging observations
of the nucleus of M~87. The integration time was 100~s per baseline, and 
we used 170 
baselines for a total on-source integration time of 5~hours. Shown are
contour plots of the model source after convolution with the
``clean beam'' (top row), the direct Fourier inversion (center row),
and the deconvolved Fourier inversion (bottom row) after 200 MEM iterations.
A 0.05\as\ region
around the central bright continuum source has been masked {\bf after}
the simulation to better display the extended stucture. Contour levels 
in all plots are 
10, 20, 30\ldots 100\% of the peak surface brightness, which is given
below each plot in units of $10^{-13}$~ergs/s/cm$^2$/\AA /\bs \label{fig6}}
\figcaption[fig7.ps]{Profile plots for the model source (top), the direct Fourier 
inversion (middle), and the restored image after 200 MEM iterations (bottom). \label{fig7}}
\figcaption[fig8.ps]{Structure of the blueshifted wing of the \ha\ emission 
(top), direct Fourier inversion (center), and reconstructed image
after 200 MEM iterations (bottom) for black hole
masses of 2, 3, 4$\cdot 10^9$\msol. 
Contour levels are as in Fig.~\ref{fig6} . \label{fig8}}
\figcaption[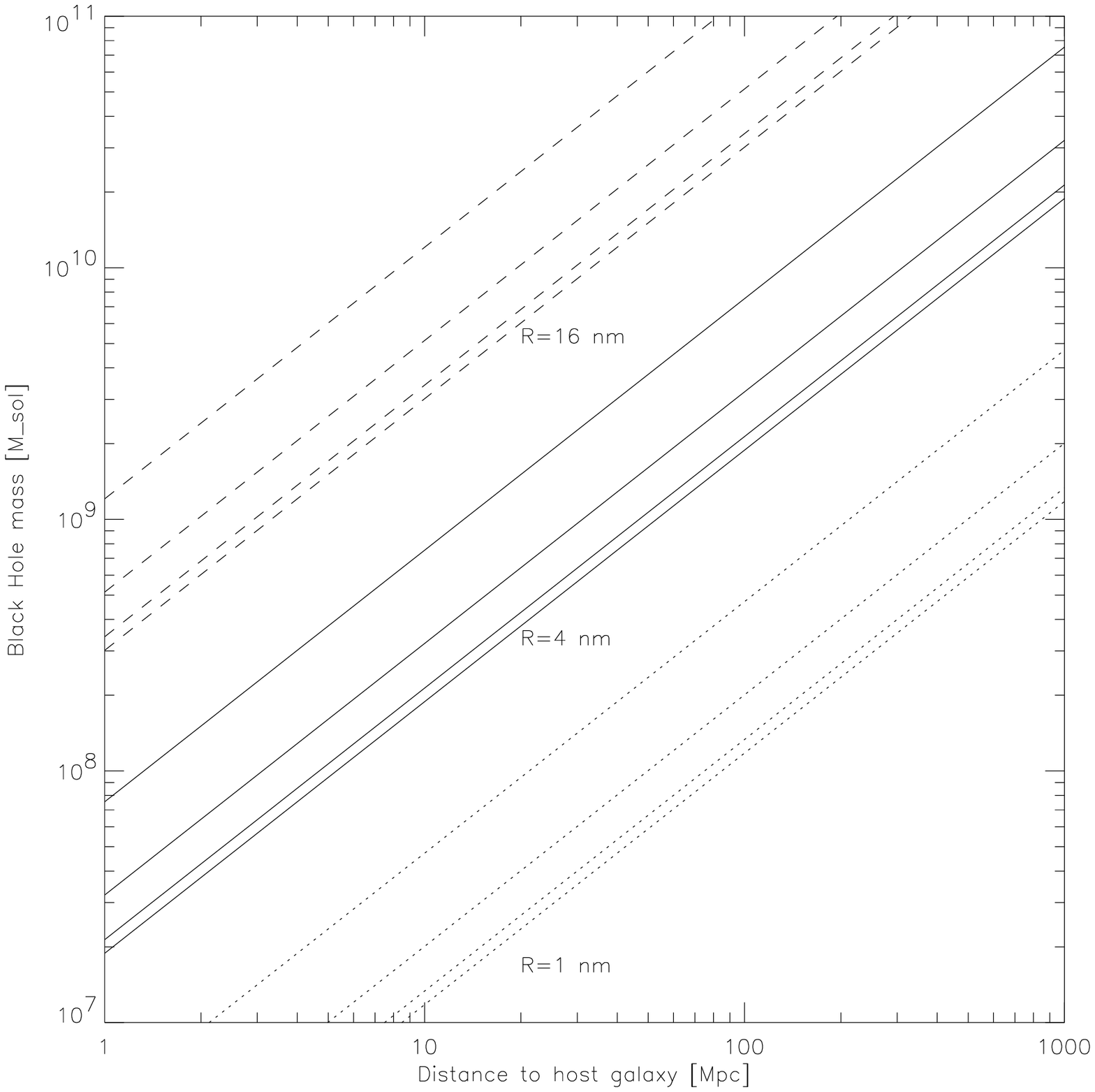]{Limiting black hole mass for the detection of a rotating gas
disk around a black hole with SIM. The Doppler shift of the line has to be at
least equal to the SIM channel width. This limit is shown
as a function of black hole mass, disk inclination, and SIM spectral
resolution. The solid lines refer to 4~nm wide SIM channels, dotted lines 
are for 1~nm channels, and dashed lines for 16~nm channel width. Four lines
per set are shown, referring to disk inclinations of $i$=0\deg\ (edge-on), 
20\deg , 40\deg , and 60\deg , the last being the
highest curve of each set. \label{fig9}}
\figcaption[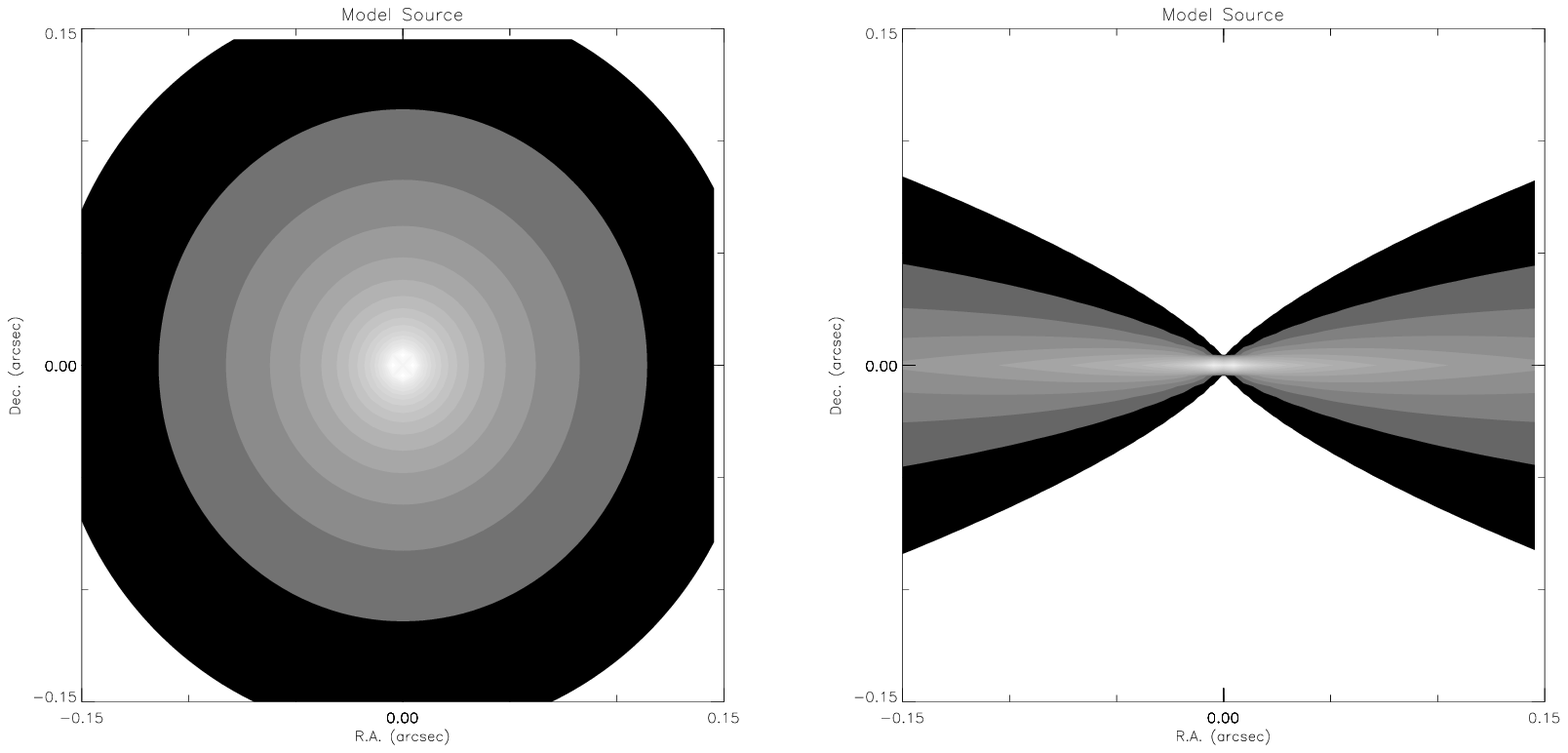]{Surface brightness distribution of our zodiacal disk model.
in face-on (left, $i=90^{\circ}$) and edge-on orientation 
(right, $i=0^{\circ}$). The grey scale is linear and identical in both plots to
emphasize the higher surface brightness in the edge-on case. The
central star has been removed for better contrast in the display. \label{fig10}}
\figcaption[fig11.ps]{Results of the \ss\ reconstruction for a \bpic -like zodiacal
disk oriented at an inclination of 0\deg\ (edge-on, left panel) and 90\deg\
(face-on, right panel). The columns in each panel are for distances 
of 100~pc (left column), 500~pc (center), and 1~kpc (right).
The rows in each panel are for the input model after convolution with the
``clean beam'' (top row), the 
direct Fourier inversion (center row) and reconstructed image
after 200 MEM iterations (bottom row). 
A 0.05\as\ diameter region around the central star has been masked out 
{\bf after} the simulation
to better show the extended, low-level structure. All panels comprise
a 0.3\as\ field of view. Contour levels are 
10,20,30$\ldots $100\% of the maximum (unmasked) surface brightness in each image 
(given below each panel in units of $10^{-15}$~ergs/s/cm$^2$/\AA /\bs ). 
The assumed SIM 
nulling efficiency is $\epsilon=10^{-4}$. Data were collected in five spectral 
channels (100~nm width between $\lambda=500$~nm and $\lambda=1000$~nm) 
and used for spectral synthesis imaging. \label{fig11}}
\figcaption[fig12.ps]{Results of the \ss\ reconstruction for 
models with various dust content in
the case of an edge-on (left) and face-on disk (right). The
columns in each panel denote (from left to right): 
C=1000, 100, 10 times the solar dust content. Field size and 
contour levels are as in Fig.~\ref{fig11}. \label{fig12}}
\figcaption[fig13.ps]{\ss\ reconstructions of a \bpic - like system 
(C=1000) at a
distance of 1~kpc. Shown are the results for various values of the
SIM nulling efficiency $\epsilon$. Again, the
edge-on case is shown in the left panel and the face-on case on the right. The
columns in each panel denote (from left to right): 
$\epsilon = 10^{-4}, 10^{-3}, 10^{-2}$. Field size and contour levels are as in
Fig.~\ref{fig11}. \label{fig13}}
\figcaption[fig14.ps]{\ss\ reconstruction of a \bpic -like zodiacal
disk (C=1000) at $d=$500~pc without (top) and with (bottom) a 20~AU radius inner
clear region. The central star has been masked {\bf after} the
simulation to better show the faint disk emission. \label{fig14}} 
\figcaption[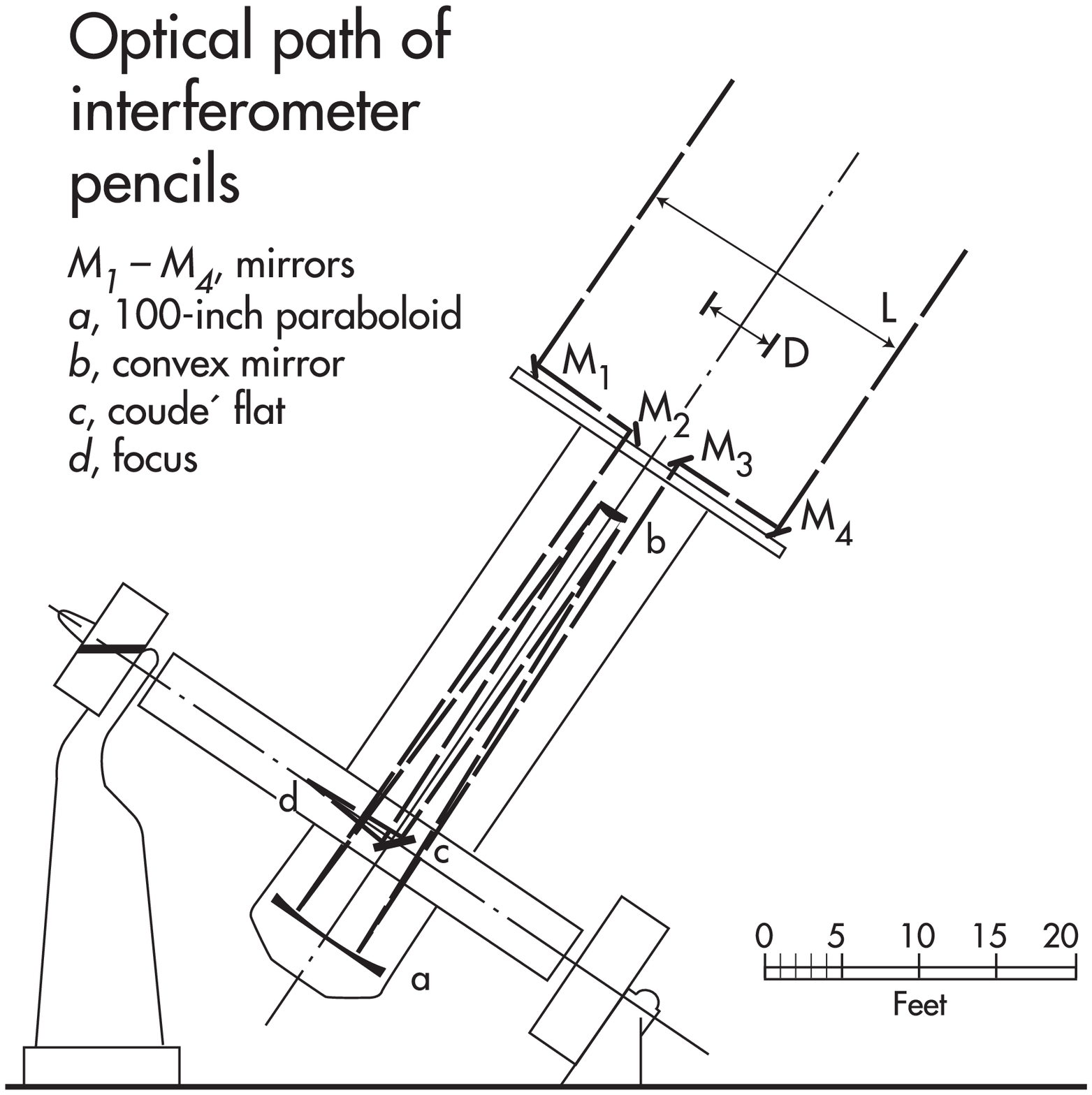]{Sketch of Michelson's stellar interferometer on the 
100-inch telescope.\label{fig15}} 
\figcaption[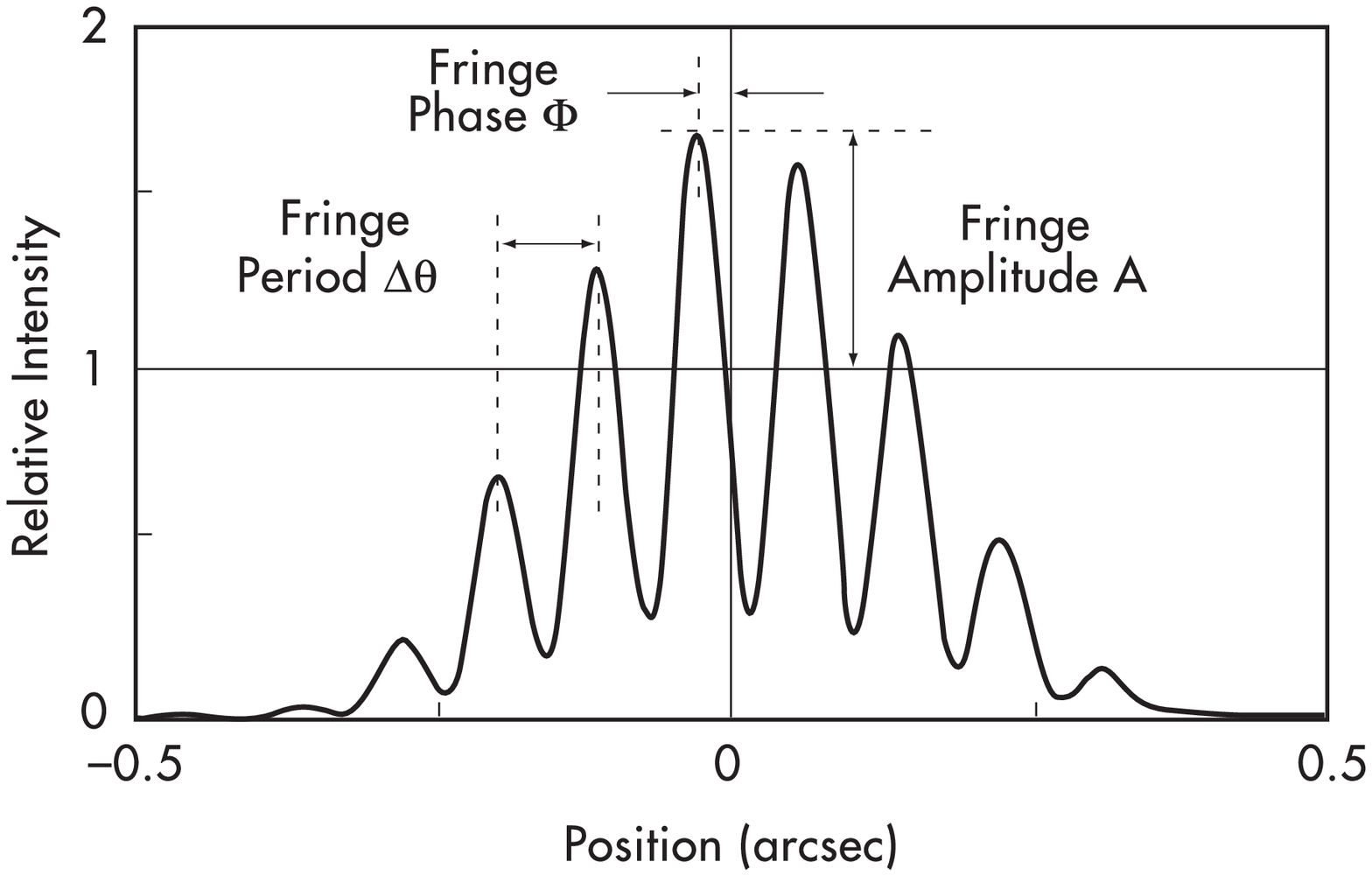]{Focal plane profile of a star in Michelson's stellar 
interferometer.\label{fig16}} 
\figcaption[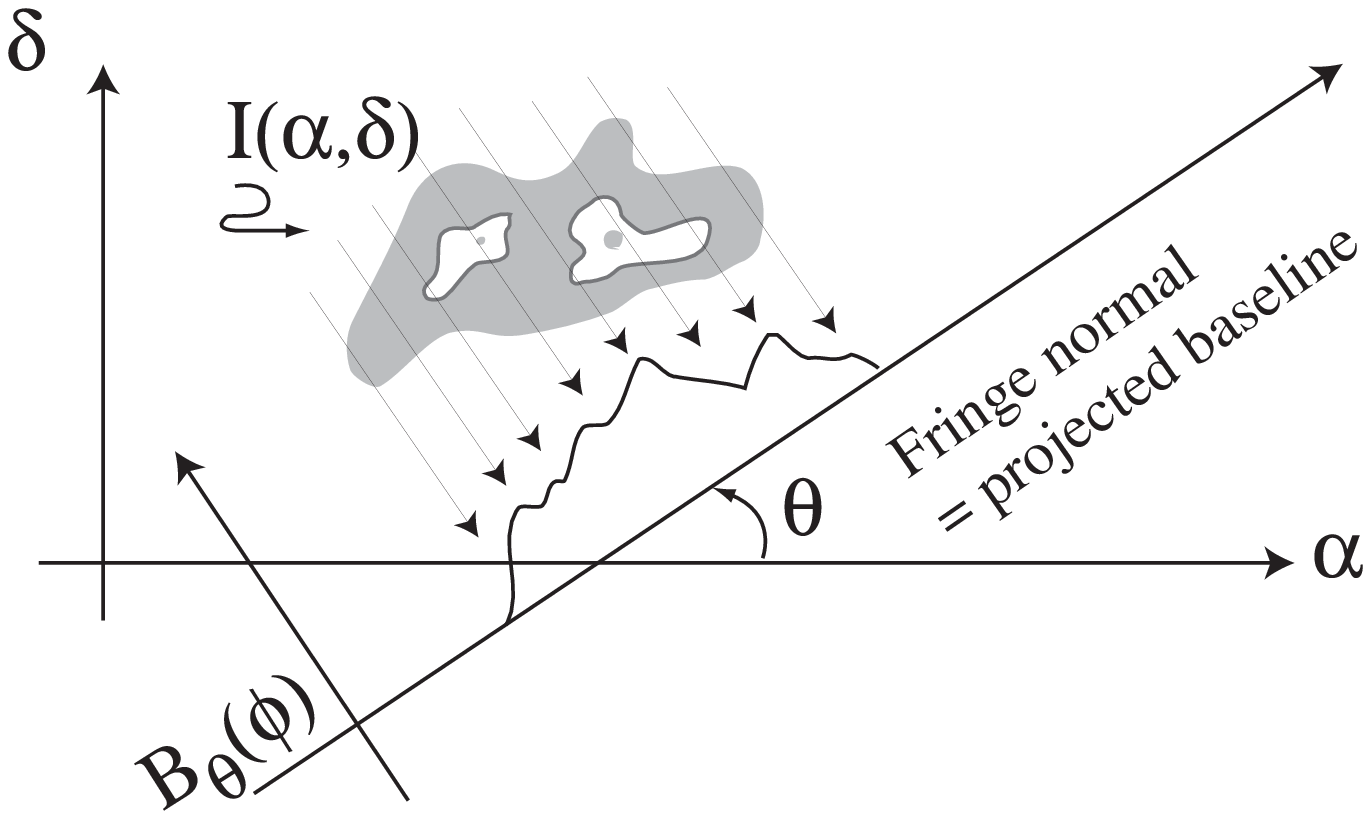]{Geometry for collapsing a 2-D source onto the fringe normal 
direction.\label{fig17} } 
\figcaption[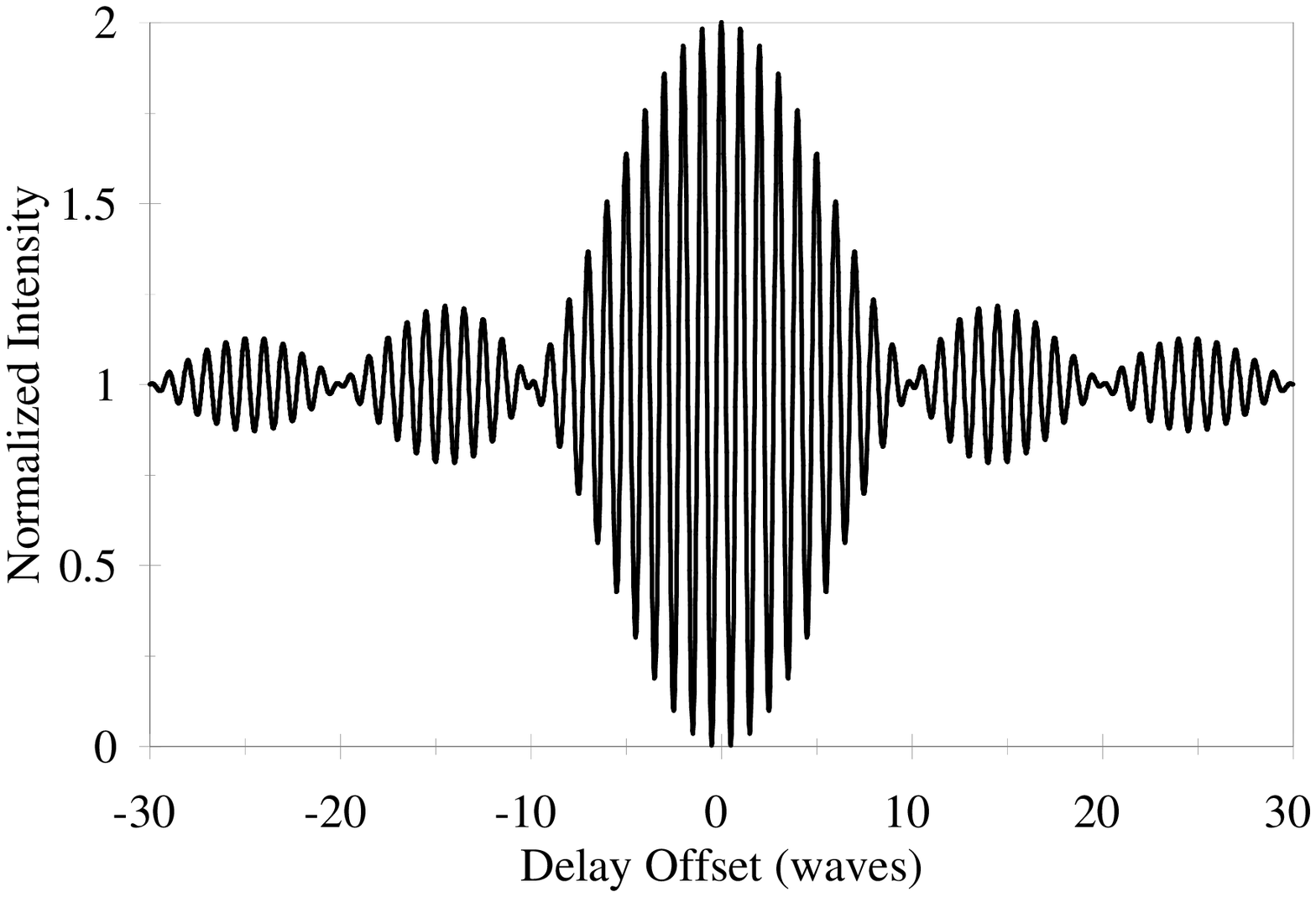]{SIM fringe pattern for a rectangular filter function.\label{fig18}} 

\newpage
\begin{deluxetable}{lcccl}
\tablenum{1}
\tablewidth{0pt}
\tablecaption{Input parameters for the M~87 model \label{tab1}}
\tablehead{
\colhead{Property} & \colhead{} & \colhead{Ref.}}         
\startdata
stellar background & $\mu_I$ = 15.0~mag/\bs & 1 \\
central point source & I = 16.1 & 1 \\
H$_{\alpha}$(+ [NII]) disk flux & $1\cdot 10^{-12}$~ergs/cm$^2$/s/\bs & 2 \\
Radial intensity profile & $I(r) = A\cdot e^{r/0.1^{\prime\prime}}$ & \\
black hole mass & $3\cdot 10^9$~\msol & 3 \\
distance & 16~Mpc & \\
inclination & 45\deg & \\
position angle & 0\deg & \\
grid resolution & 0.005\as &
\enddata
\tablecomments{References: 1) \cite{lau92}; 2) \cite{har94}, assuming
a line ratio of [SII]/(\ha +[NII]) = 4;  3) \cite{mar97}. }
\end{deluxetable}

\clearpage
\plotone{fig4.ps}

\centerline{Fig. 4}

\newpage
\plotone{fig5.ps}

\centerline{Fig. 5}

\newpage
\plotone{fig9.ps}

\centerline{Fig. 9}

\newpage
\plotone{fig10.ps}

\centerline{Fig. 10}

\newpage
\plotone{fig15.ps}

\centerline{Fig. 15}

\newpage
\plotone{fig16.ps}

\centerline{Fig. 16}

\newpage
\plotone{fig17.ps}

\centerline{Fig. 17}

\newpage
\plotone{fig18.ps}

\vfill
\centerline{Fig. 18}
\end{document}